\documentclass[tightenlines,superscriptaddress,eqsecnum,floats,nofootinbib,showpacs]{revtex4}

\usepackage{amssymb}
\usepackage{stmaryrd}
\usepackage{amsmath}
\usepackage{amsfonts}
\usepackage{mathrsfs}
\usepackage{CJK}
\usepackage{amsmath,amssymb,amsfonts}
\usepackage{graphicx}

\def\be{\begin{equation}}
\def\ee{\end{equation}}
\def\ba{\begin{eqnarray}}
\def\ea{\end{eqnarray}}

\newcommand{\abs}[1]{{\left|{#1}\right|}} 

\begin{document}


\title{A possible mechanism for over luminous type Ia supernovae explosions inspired by dark matter}

\author{Zhenzhen Jing}
\affiliation{School of Science, South China University of
Technology, Guangzhou 510641, P.R. China}

\author{ Xiangdong Zhang\footnote{Corresponding author: scxdzhang@scut.edu.cn}}
\affiliation{School of Science, South China University of
Technology, Guangzhou 510641, P.R. China}

\author {Dehua Wen }
\affiliation{School of Science, South China University of
Technology, Guangzhou 510641, P.R. China}

\date{\today}


\begin{abstract}

Dark matter is believed to be a major component of our universe. In this paper we propose a new mechanism based on
dark matter inspired super-Chandrasekhar mass white dwarf to explain the recent observation of super luminous type Ia supernovae explosions.
 Our calculation shows when a white dwarf accretes enough dark matter, due to the Pauli exclusive principle between fermionic dark matter particles,
 the mass of corresponding  dark white dwarf (which means the white dwarf mixed with dark matters) can significantly exceeds the Chandrasekhar limit.
 Moreover, we investigate some  physical observable quantities, such as the redshift and moment of inertia of the dark white dwarf and found that
  these quantities are sensitive to the dark matter particle's distributions and thus can be potentially used to probe the relevant
  information of dark matter particles in the future.

\end{abstract}


\maketitle

\section{Introduction}

Type Ia supernovae(SNe Ia) are of vital importance as luminosity distance indicators for measuring the expansion history of the
 universe \cite{Riess1998,Perlmutter1999}. SNe Ia are widely believed to arise from the drastic thermonuclear explosion of white dwarf, when its mass comes near
 to the Chandrasekhar limit of $1.44M_\odot$, here $M_\odot$ is the solar mass. However, some documented examples of overluminous SNe Ia explosions e.g., SN2003fg,
  SN2006gz, SN2007if and SN2009dc \cite{Howell06,Scalzo10} has been reported recently. One natural explanation of such peculiar phenomenon is to suppose the mass
   of progenitor star of these SNe Ia potentially exceeding $1.44M_\odot$. Thus, in order to naturally generate super-Chandrasekhar mass white dwarf, several
    models have been proposed, such as highly magnetized white dwarf \cite{Mukhopadhyay12,Das12,Das13,Wen2014}, rotating white dwarf \cite{Hachisu2012,Yoon2005},
     electronic charged white dwarf \cite{liu2014}, white dwarf in modified general relativity \cite{Das2015,Das20152,Jing2016}, et al. However, none of these
      models have been directly verified by concrete experimental data. Therefore exploring other physical viable mechanism is not only meaningful at
       theoretical level but also can serve as the guideline for the future experiment to unveil such mysterious phenomenon.

On the other hand, it has been widely accepted that nearly a quarter of the mass of our universe is in the form of dark matter(DM). Observations of the flatness
 of galactic rotation curves, measurements of the cosmic microwave background, and baryon acoustic oscillations\cite{K0812,G0907,M1001,G2005,G2013,p2015}
  proof the existence of DM. Nevertheless, the nature of dark matter is
   still elusive. On the theoretical side, one of the main streams to explain the phenomenon of DM is to suppose the existence of one or some kinds of unveiled
    DM particles. Along this direction, people propose many possible candidates of DM particles, such as axion particle, sterile neutrino, supersymmetric particles
    and so on.  Nowadays, many apparatuses are set up to search these mysterious particles. However, since the interactions between DM particles and normal matters
     are extremely weak, until now, no completely convincing evidences have been found. Therefore, besides these directly experimental hunting, even finding some
     indirect evidences or influences of DM particle will be greatly appreciated.  Note that like normal matter, DM particles also have gravitational interaction.
     This important feature of DM particles hint us that some interesting consequences might be found in the  massive objects. In the astrophysics,
     these massive objects such as white dwarfs, neutron stars are widely distributed in our universe. In
     Refs.\cite{Laura2015,Payel2015,Leung2013,Fredrik2009,paolo2011,Ang2012} the authors study the effect of DM on compact objects which formed by DM admixed with ordinary matter made of neutron star matter, white dwarf matter or quark star material, in order to investigate the properties of DM,
     such as their spin, mass and interaction. In \cite{Leung2013}, they assume the equilibrium structures of white dwarfs with DM cores by
     non-selfannihilating DM particles in general relativistic two-fluid formulation. Inspired by the previous studies and to explain the phenomenon
     of overluminous SNe Ia explosions, we propose a new model, assuming DM particles are distributed in white dwarf to support the super-Chandrasekhar
     mass white dwarf. In this work, we will focused on the fermionic DM particles with light mass, for example, the sterile neutrino are perfectly well suited
     in our model. Since the interaction between DM and normal matter are extremely weak, as a first order approximation, we can simply assume the DM particles
     are in the form of free fermion gas at zero temperature\cite{zheng2016}\cite{Narain2006}. Because of Pauli inclusive principle, there is a repulsive force
      associated with this free dark matter gas, and this in turn provides a required repulsive force to stiffen the equation of state (EOS), and then
       lift the mass of the white dwarf to exceed the
      Chandrasekhar limit.   Through above procedures,
       we study the structure of white dwarf with free DM gas, and use it to explain the over-lumimous SNe Ia. Moreover, In order to find some possible
        observation signals of our model, the red shift and the moment of inertia of our super-Chandrasekhar mass white dwarf is also been studied.

This paper is organized as follows: After a short introduction, we
present the basic structure formulation of the white dwarf mixed with DM in
Sec. II , where two subsections are included£¬ the exact solution of the dark white dwarf is obtained in subsection A, while the EOS of dark white dwarf and  the corresponding numerical results are shown in subsection B, General relativity effects of the dark white dwarf are discussed in subsection C. The redshift and the moment of inertia of dark compact star is shown in Sec. III, Conclusions and outlooks are given in the last section.

\section{Hydrostatic Equilibrium Equation of dark matter mixed White Dwarf}

 For a static and spherical symmetric
dark white dwarf, in the Newtonian framework, the hydrostatic equilibrium
equation can be written as
\begin{equation}\label{dp}
\frac{dp}{dr}=-\frac{Gm(\rho_m+\rho_{dm})}{r^2},
\end{equation}
where $\rho_m$ and $\rho_{dm}$ are mass density of normal matter and  dark matter respectively inside the white dwarf;
the stellar mass $m$ inside the radius $r$ is related with the matter density through
\begin{equation}\label{dm}
\frac{dm}{dr}=4\pi r^2(\rho_m+\rho_{dm}).
\end{equation}
Note that there is no prior principle to determine the
distribution of DM particle inside the white dwarf, in order to
calculate the structure of dark white dwarf, we need to given such
distribution of DM by hand.  One of the most simplest choice is to
assume the dark matter number density of dark white dwarf $n_{dm}$
proportional to the electron number density $n_e$, that is,
\begin{equation}\label{ndm}
n_{dm}=\alpha\times{n_e}=\alpha\times\frac{\rho_m}{m_{n}\times{\mu_{e}}},
\rho_{dm}=n_{dm}\times{m_{dm}},
\end{equation} where $\alpha$ is a dimensionless proportionality coefficient, $m_{n}$ and $m_{dm}$ is the mass of a nucleon and dark matter particle, respectively. $\mu_e$ is the ratio of nucleon numbers to electron numbers. Other distributions are of course allowed in
principle, however, this will only change the result quantitatively, the whole qualitative picture remains valid.

Applying the above hydrostatic equilibrium equation to polytropic newtonian star we can get an exact solution in the following subsection A.

\subsection{exact solution for polytropic equation of state}

Note that the equation of state(EOS) of a free fermion gas can be well approximated to a polytropic EOS. So we first use a polytropic EOS in order to make the calculation expediently and consider the more realistic EOS in the next subsection. Thus the total equation of state of dark white dwarf is
\ba
p=\kappa\rho_{m}^{1+\frac{1}{n}}\times(1+\alpha^{1+{\frac{1}{n}}}),\label{dp3}
\ea where $n$ is an integer.
The equilibrium equation (\ref{dp}) can be reduced to
\ba
\frac{dp}{dr}=-\frac{Gm(\rho_m+\rho_{dm})}{r^2}=-\frac{Gm\rho_m}{r^2}\left[1+\alpha\eta\right], \label{dp2}
\ea
\ba
\frac{dm}{dr}=4\pi{r^2}\rho_m(1+\alpha\eta),\label{dm2}
\ea where
$\eta=\frac{m_{dm}}{\mu_{e}m_n}$
is a dimensionless constant depend on the mass of DM particle. The Eqs.(\ref{dp2}) and (\ref{dm2}) can be unified as
\ba
\frac{1}{r^2}\frac{d}{dr}\left(\frac{r^2dp}{\rho_m dr}\right)=-4\pi
G\rho_m\left[1+\alpha\eta\right]^2.\label{dp1}
 \ea
  Introduce $\rho=\rho_{mc}\theta^n, r=a\xi$, where $\rho_{mc}$ is the central density of the white
  dwarf,  then   Eq. (\ref{dp3}) can be written as
  \ba
  p=\kappa\rho_{mc}^{(1+1/n)}\theta^{(1+n)}(1+\alpha^{(1+1/n)}),
  \ea
where $a$ has the dimension of length and is  defined by \ba
a=\left[\frac{(n+1)\kappa\rho_{mc}^{\frac{1}{n}-1}(1+\alpha^{(1+\frac{1}{n})})}{4\pi
G\left(1+\alpha\eta\right)^2}\right]^{\frac{1}{2}}. \ea Through the transform, Eq.(\ref{dp}) is reduced to the standard form of Lane-Emden
equation\cite{Weinberg72}
 \ba
\frac{d}{d\xi}\left(\xi^2\frac{d\theta}{d\xi}\right)+\theta^n\xi^2=0\label{LE}.
\ea
 Add the boundary conditions
 $\theta(\xi=0)=1, \frac{d\theta}{d\xi}\Big|_{\xi=0}=0$,
Eq.(\ref{LE}) can be solved analytically. The radius of dark white
dwarf is given by $R=a\xi_1$, where $\xi_1$ is the first zero
point of $\theta$ function. The mass of dark  white dwarf can be
obtained\ba M=4\pi
a^3\rho_{mc}(1+\alpha\eta)\xi_1^2\abs{\theta'(\xi_1)}. \ea
Consider the index $\gamma=\frac43$ (or equivalently $n=3$), we
found the mass of dark white dwarf becomes independent with
$\rho_{mc}$, and thus corresponds with the maximal mass of dark
white dwarf. By comparing these results, we obtain the relation
between dark white dwarf and ordinary white dwarf, that is
  \ba
R=\frac{R_{Ch}\left(1+\alpha^{\frac{4}{3}}\right)^\frac12}{1+\alpha\eta},\quad\quad  M=\frac{M_{Ch}\left(1+\alpha^{\frac{4}{3}}\right)^\frac32}{(1+\alpha\eta)^2}\label{MR}
\ea
 where $R_{Ch}$ and $M_{Ch}$ are the Chandrasekhar radius and Chandrasekhar mass limit for ordinary white dwarf, respectively. Finally,
 it is worth noting that when $\alpha=0$ the Chandrasekhar mass limit of ordinary white dwarf will be recovered perfectly.

\subsection{EOS of dark  white dwarf and numerical  results}
 We have been assumed that the repulsive force of compact star which balance gravity is formed by free fermionic dark matter gas and free electron gas from normal matter, the mixed EOS for free fermionic gas at zero temperature can be obtained via the explicit expressions for maximal momentum of fermion and pressure,
\begin{equation}\label{kF}
k_{_F}=\hbar(3\pi^2n_i)^{1/3},
\end{equation}

\begin{equation}\label{p2}
p=\frac{8\pi c}{3(2\pi\hbar)^3}\int_0^{k_F}\frac{k^2}{(k^2+m_i^2c^2)^{1/2}}k^2dk
\end{equation}
where $i=1$, for electron; $i=2$, for DM particles , $k$ represents the momentum of fermion and $k_F$ is the
maximum momentum determined by mass density. In the above equations, $n_e$ is the number density of electron, $n_{dm}$ is DM number density. It is easy to see that even
in this simple case, the equation of state still can not be
attributed to polytropic type, and thus become difficult to be
solved analytically. In view of these facts,  the equations
(\ref{dp})-(\ref{dm}) should be treated by numerical calculation.

\begin{figure}\label{MR0}
\centering
\includegraphics [width=1\textwidth]{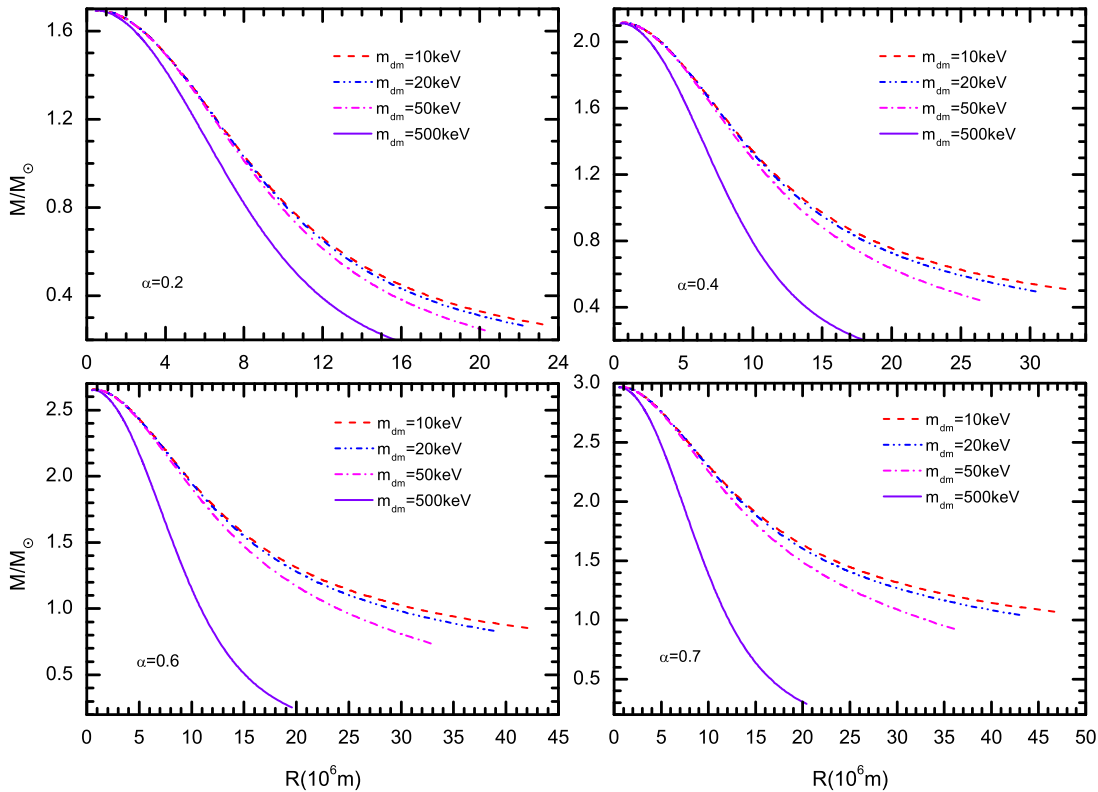}
\caption{\label{mr0} The mass-radius relations of the dark white dwarf
 for different $\alpha$ and different
$m_{dm}$, where the central density ranges from
$5\times10^8kg/m^3$ to $4.3\times10^{14}kg/m^3$. }
\end{figure}

\begin{figure}\label{MR1}
\centering
\includegraphics [width=0.7\textwidth]{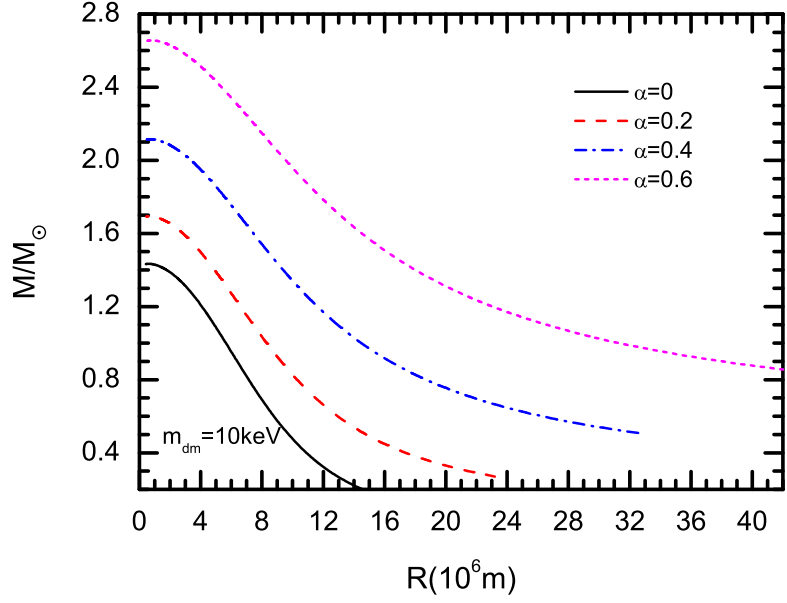}
\caption{\label{mr1}   The mass-radius relations of the dark white
dwarf for  different  value of $\alpha$ with fixed
$m_{dm}(=10keV)$. The black line is the curve of $\alpha=0$ which
matches the famous Chandrasekhar result $M_{max}\sim1.4M_\odot$.}
\end{figure}

\begin{figure}\label{PRHO}
\centering
\includegraphics [width=0.7\textwidth]{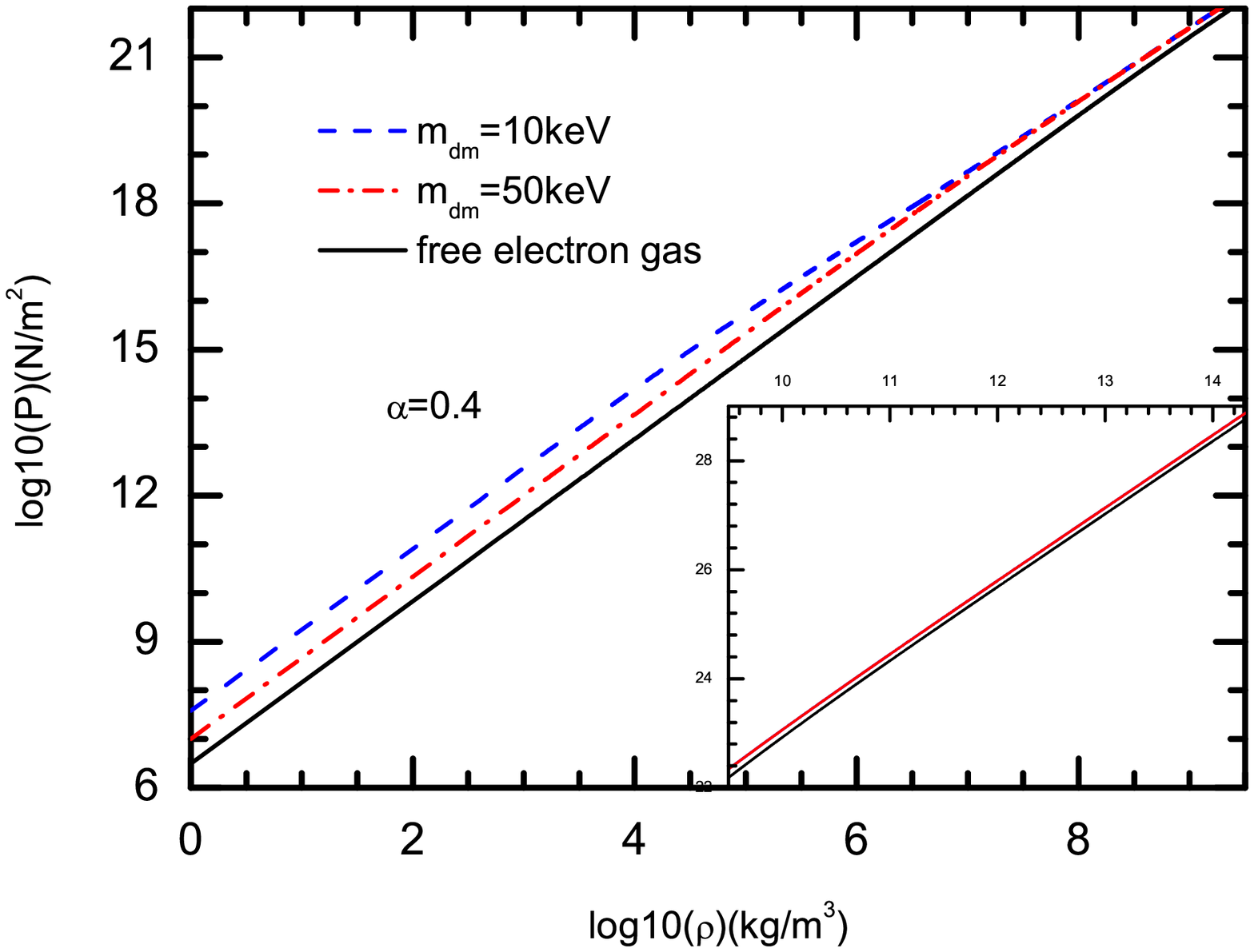}
\caption{\label{prho} EOS of the  dark white dwarf for
$m_{dm}=10keV$ and $m_{dm}=50keV$ with $\alpha=0.4$, and the black
line is the EOS of free electron gas. The inset shows the EoS in
high density region. }
\end{figure}

Combining the above EOS with the Newtonian equilibrium equations
(\ref{dp})-(\ref{dm}) and the boundary conditions: $m=0$ at $r=0$
and $p=0$ at $r=R$, for a given mass of DM particle, we
can obtain the mass-radius relation for dark white dwarf
numerically. The numerical results for dark white
dwarf are presented in Fig. \ref{mr0} and Fig.\ref{mr1}. Fig. \ref{mr0} shows the mass-radius relations for four typical masses
of DM particles ranging from $m_{dm}=10 keV$ to
$m_{dm}=500 keV$ with different distribution
$\alpha$($\alpha=0.2-0.7)$. It is shown that the maximal mass of
dark white dwarf sensitive to the value of $\alpha$, while
unsensitive to the mass of DM particle. However with the central
density decreasing, the curves get more and more deviate which
indicate that the  mass of DM particle mainly affects the radius
of dark white dwarf at low density region.

In order to show the impacts of number density of DM on
the white dwarf intuitively,  and also to compare them with that
of the ordinary white dwarf, the mass-radius relation for
different $\alpha$ with the same mass of DM particle
$m_{dm}=10keV$ is presented in Fig. \ref{mr1}. It shows that the
mass of dark white dwarf is larger than ordinary white dwarf when
$\alpha$ increases. For the $\alpha$ takes a range from $0.2$ to
$0.6$, the corresponding maximum mass of dark white dwarf varies
from $1.7M_\odot$ to $2.7M_\odot$. This in turn provides a perfect
mechanism to explain the entire range of the observed peculiar
overluminous super-SNeIa date.

Fig. \ref{prho} shows that the EoSes of  dark white dwarfs for
$m_{dm}=10keV$ and $m_{dm}=50keV$ with $\alpha=0.4$ and the EoS of
free electron gas. We can see that these EOS only difference at
the lower density region, while at the higher density region, the
difference between these EOSes become negligible. Since the
maximal mass of compact star is mainly determined by its higher
density part of EOS, it is easy to understand that why the maximal
mass of dark white dwarf unsensitive to the mass of DM particle.

\subsection{General relativity effects of the dark white dwarf}
From the numerical calculation, the mass of dark  white dwarf can
reach up to $2.9M_\odot$ when $\alpha=0.7$. The compact star is so
massive that people might worry about using Newtonian hydrostatic
equilibrium framework. Should we consider the general relativity
effect? To this aim, we calculate the compactness parameter of
dark white dwarf $\chi=\frac{GM}{c^2R}$ firstly. For $\alpha=0.4$,
the detailed calculation shows the compactness of the maximum-mass
dark white dwarf is about
 \ba
\chi=\frac{GM}{c^2R}\thickapprox 4.5\times10^{-4}, \ea
which means we can
safely ignore the effect of general relativity on the dark white
dwarf.

\section{ redshift and  moment of inertia of dark white dwarf}
\begin{figure}
\centering
\includegraphics [width=0.7\textwidth]{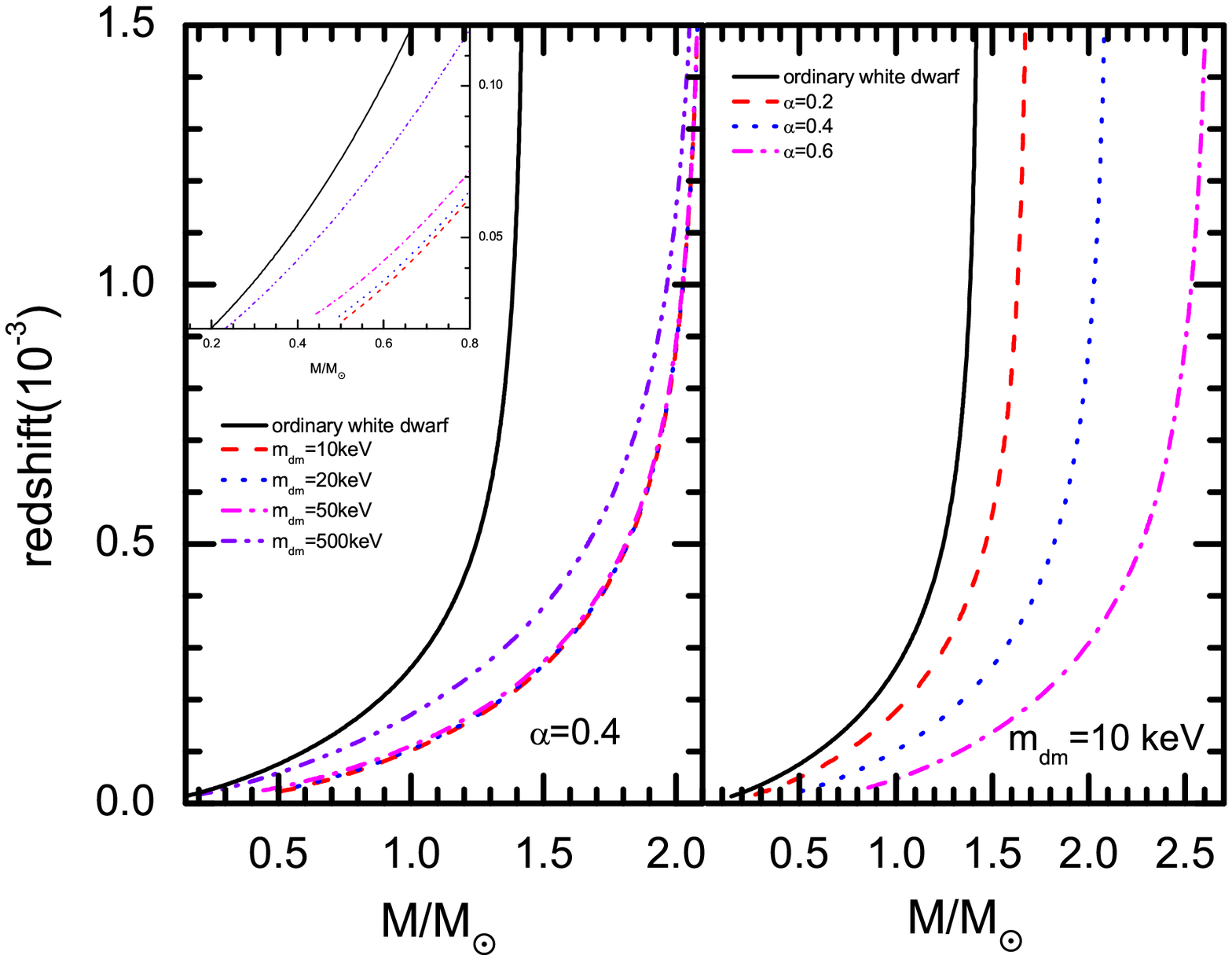}
\caption{\label{R} The redshift  as a function of the stellar mass
for different DM particle mass with fixed $\alpha (=0.4)$(left)
and for different $\alpha$ with fixed $m_{dm}$ (=10keV)(right)}
\end{figure}
\begin{figure}
\centering
\includegraphics [width=0.7\textwidth]{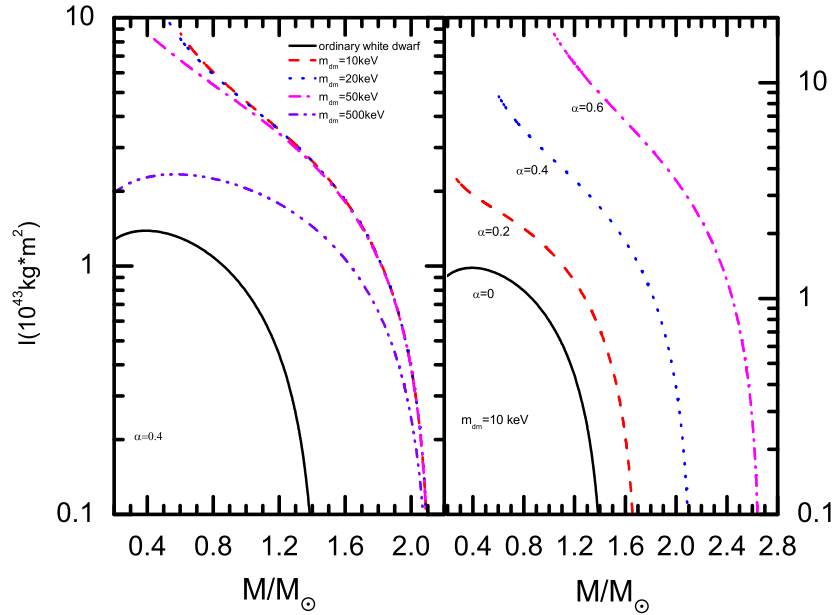}
\caption{\label{R-alpha} The moment of inertia   as a function of
the stellar mass for different DM particle mass with fixed $\alpha
(=0.4)$(left) and for different $\alpha$ with fixed $m_{dm}$
(=10keV)(right) }
\end{figure}

Except mass and radius, there are several other physical
quantities which characterize the dark white dwarf,  such as
redshift and moment of inertia. From above discussion, the radius
of small mass dark white dwarf sensitive to the mass of DM particle. Thus, the DM particle also have effects
on the redshift and moment of inertia. Therefore the
investigations on the redshift and moment of inertia of dark white
dwarf in turn can shed some lights on the study of property of DM
particles.

For a static spherically symmetric compact star, the redshift of
the star can be written as
$z=\frac{1}{\sqrt{1-\frac{2GM}{c^2R}}}-1$, and in the framework of
Newtonian dynamics, The moment of inertia reads \ba
I=\frac{8\pi}{3}\int_{0}^{R}drr^4{\rho}. \ea Using these two
expressions, the numerical results for redshift and moment of
inertia is presented in Fig.{\ref{R}} and Fig.{\ref{R-alpha}}.
Fig.{\ref{R}} shows   the redshift as a function of mass, where
the left picture shows  the redshift as functions of the total
star mass for different DM particle mass for a fixed
$\alpha(=0.4)$. The redshift of dark white dwarf  is lower than
the ordinary white dwarf when the mass below $1.44M_\odot$, and
the inset picture clearly shows that at lower stellar masses,
higher DM particle's corresponds to larger redshift, thus the
redshift of dark white dwarf reflects the information of
 DM particle's mass. The right part of   Fig.{\ref{R}} shows
the redshift-mass relation for different $\alpha$ with
$m_{dm}=10keV$, the result shows that the redshift is inversely
related to $\alpha$ for a fixed mass of DM particle.

The relationship between moment of inertia and mass is presented
in Fig.{\ref{R-alpha}}.  The black line is the ordinary white
dwarf with maximum mass of $1.4M_{\odot}$. It is shown that for a
fixed mass of DM particle $m_{dm}=10keV$, the moment of inertia of
dark white dwarf increasing as the parameter $\alpha$ arising.
While for a fixed $\alpha=0.4$, the mass of DM particle can be
distinguished by its moment of inertia of dark white dwarf. Hence
the future observational data of moment of inertia have potential
opportunity to reveal the relevant information of DM particle
inside the dark white dwarf.

\section{Conclusions}
In this paper, we propose a possible mechanism to explain some
peculiar type Ia supernova explosion, since our universe has
abundant DM,  we suggest to consider the possibility of
white dwarf accretes DM. We calculate the structure of
the corresponding dark white dwarf. Our calculations show that
when a white dwarf accretes enough DM, the mass of that dark white
dwarf can be uplifted significantly to exceed the Chandrasekhar
limit, and hence can naturally explain the observed peculiar Type
Ia supernova explosion. Particularly, by employing a
representative choice of DM distribution, we obtain an analytic
solution for the stellar structure, as shown in Eq.  ({\ref{MR}}).
 Note that the radius of lower mass dark
white dwarf is sensitive to DM particle's mass, we further investigate
other physical observable quantities, such as redshift and moment
of inertia of dark white dwarf. Our result suggests that the
redshift and moment of inertia of dark white dwarf is
significantly different with ordinary white dwarf, and this in
turn provides a new way to detect relevant information of DM
particles though astrophysics observations in the future. We hope
this can open a new windows to shed some lights on hunting DM
particles.

\begin{acknowledgements}
This work is supported by NSFC ( Nos.10947023, 11275073 and
11305063) and the Fundamental Research Funds for the Central
University of China. This
project is sponsored by SRF for ROCS and SEM and has made use of NASA's
Astrophysics Data System.

\end{acknowledgements}


\end{document}